\documentclass[pdflatex,sn-nature,Numbered]{sn-jnl}

\usepackage{graphicx}%
\usepackage{multirow}%
\usepackage{amsmath,amssymb,amsfonts}%
\usepackage{amsthm}%
\usepackage{mathrsfs}%
\usepackage[title]{appendix}%
\usepackage{xcolor}%
\usepackage{textcomp}%
\usepackage{manyfoot}%
\usepackage{booktabs}%
\usepackage{algorithm}%
\usepackage{algorithmicx}%
\usepackage{algpseudocode}%
\usepackage{listings}%
\usepackage{ulem}
\usepackage{comment}
\usepackage{lineno}

\theoremstyle{thmstyleone}%
\theoremstyle{thmstyletwo}%

\theoremstyle{thmstylethree}%

\newcommand{\uud}{$\uparrow \uparrow \downarrow$}
\newcommand{\ud}{$\uparrow \downarrow$}

\begin{document}

\title[Article Title]{Atomically-sharp magnetic soliton in the square-net lattice EuRhAl$_{4}$Si${_{2}}$}

\author[1,2]{\fnm{Kevin} \sur{Allen}}
\author[3]{\fnm{Juba} \sur{Bouaziz}}
\author[1,2]{\fnm{Yichen} \sur{Zhang}}
\author[4,5]{\fnm{Kai \sur{Du}}}
\author[1,2]{\fnm{Sanu} \sur{Mishra}}
\author[3]{\fnm{Gustav} \sur{Bihlmayer}}
\author[6]{\fnm{Yiqing} \sur{Hao}}
\author[7]{\fnm{Victor} \sur{Ukleev}}
\author[7]{\fnm{Chen} \sur{Luo}}
\author[7]{\fnm{Florin} \sur{Radu}}
\author[1,2]{\fnm{Yuxiang} \sur{Gao}}
\author[8]{\fnm{Marta} \sur{Zonno}}
\author[8]{\fnm{Sergey} \sur{Gorovikov}}
\author[9]{\fnm{Christopher} \sur{Lane}}
\author[9]{\fnm{Jian-Xin} \sur{Zhu}}
\author[6]{\fnm{Huibo} \sur{Cao}}
\author[4,5]{\fnm{Sang-Wook} \sur{Cheong}}
\author[1,2]{\fnm{Ming} \sur{Yi}}
\author[3,10]{\fnm{Stefan} \sur{Bl{\"u}gel}}
\author*[1,2]{\fnm{Emilia} \sur{Morosan}}\email{emorosan@rice.edu}

\affil[1]{\orgdiv{Department of Physics and Astronomy}, \orgname{Rice University}, \orgaddress{\city{Houston}, \postcode{77005}, \state{TX}, \country{USA}}}

\affil[2]{\orgdiv{Rice Center for Quantum Materials (RCQM)}, \orgname{Rice University}, \orgaddress{\city{Houston}, \postcode{77005}, \state{TX}, \country{USA}}}

\affil[3]{\orgdiv{Peter Gr{\"u}nberg Institut}, \orgname{Forschungzentrum J{\"u}lich \& JARA}, \orgaddress{\city{J{\"u}lich}, \postcode{D-52425}, \country{Germany}}}

\affil[4]{\orgdiv{Department of Physics and Astronomy}, \orgname{Rutgers University}, \orgaddress{\city{Piscataway}, \postcode{08854}, \state{NJ}, \country{USA}}}

\affil[5]{\orgdiv{Keck Center for Quantum Magnetism}, \orgname{Rutgers University}, \orgaddress{\city{Piscataway}, \postcode{08854}, \state{NJ}, \country{USA}}}

\affil[6]{\orgdiv{Neutron Scattering Division}, \orgname{Oak Ridge National Laboratory}, \orgaddress{\city{Oak Ridge}, \postcode{37831}, \state{TN}, \country{USA}}}

\affil[7]{\orgdiv{Helmholtz-Zentrum Berlin f{\"u}r Materialien und Energie}, \orgaddress{\street{Albert-Einstein-Str. 19}, \city{Berlin}, \postcode{D-12489},  \country{Germany}}}

\affil[8]{\orgdiv{Canadian Light Source, Inc.},\orgaddress{\street{44 Innovation Boulevard}, \city{Saskatoon}, \postcode{S7N 2V3}, \state{SK}, \country{Canada}}}

\affil[9]{\orgdiv{Theoretical Division, Los Alamos National Laboratory},\orgaddress{\city{~Los Alamos}, \postcode{87545}, \state{NM}, \country{USA}}}

\affil[10]{\orgdiv{Institute for Theoretical Physics, RWTH Aachen University},\orgaddress{\city{~Aachen}, \postcode{D-52074}, \country{Germany}}}

\abstract{
Topological spin textures are hallmark manifestations of competing interactions in magnetic matter. Their effective description by nonlinear field theories reflects an energetic frustration that destabilizes uniform order while selecting finite-size, topologically nontrivial configurations as stationary states. Among the most extreme realizations are atomically-sharp domain wall excitations, namely one-dimensional (1D) magnetic solitons, which represent the ultimate scaling limit of magnetic textures. Such solitons may emerge in magnetic systems where effective exchange interactions compete directly with uniaxial magnetic anisotropy. Here we show that the square-net rare earth compound EuRhAl$_{4}$Si$_{2}$ realizes a very susceptible regime  where the magnetic anisotropy competes with highly frustrated exchange interactions stabilizing a rare ferrimagnetic $\uparrow\uparrow\downarrow$ state that, under applied magnetic field, supports the formation of atomically-sharp soliton defects. We confirm the bulk response of the 1D magnetic solitons via magnetization and electrical transport measurements. We establish both the zero- and in-field $\uparrow\uparrow\downarrow$ order via neutron diffraction, while magnetic force microscopy visualizes its real-space evolution into a stripe-like array. To elucidate the microscopic origin of the soliton, we relate the Ruderman–Kittel–Kasuya–Yosida (RKKY)-driven exchange interactions and the magnetic anisotropy through density functional theory, and we construct an effective 1D $J_{1}$--$J_{2}$--$K$ model whose atomistic spin dynamics simulations reproduce the observed soliton states as a function of external field. Our results demonstrate that EuRhAl$_{4}$Si$_{2}$ hosts atomically-sharp, field-driven 1D magnetic solitons, providing a new platform for studying 1D topological excitations at the atomic length scale.}

\maketitle
\section*{Main}\label{sec1}
The delicate competition between magnetic interactions has emerged as a central principle that governs the physics of topological magnetic textures. The interplay between Heisenberg exchange and Dyzaloshinskii-Moriya interaction (DMI) in noncentrosymmetric magnets is a prominent such example. This interplay stabilizes two-dimensional (2D) topological solitons known as chiral skyrmions and extended micromagnetic skyrmion lattices \cite{ muhlbauer2009skyrmion, tokura2020magnetic}, and ultimately establishes the field of skyrmionics. In centrosymmetric crystals, replacing the competition between exchange interaction and DMI with appropriate competition between ferro- and antiferro-magnetic interactions can lead to skyrmions and skyrmion lattices even in achiral crystals, with the skyrmions an order of magnitude smaller in size (1 - 10 nm)\cite{hayami2021square,hayami2022rectangular,bouaziz2022fermi} than in non-centrosymmetric systems. Rare earth magnets containing Eu or Gd epitomize this behavior when their frustrated exchange introduced by the oscillatory Ruderman-Kittel-Kasuya-Yosida (RKKY) interaction can overcome the magnetic anisotropy due the lack of any strong single-ion anisotropy. In many such systems, the frustrated exchange energy landscape favors Yoshimori spirals \cite{yoshimori1959new}, a  type of spin spiral whose symmetry-degenerate variants combine under magnetic field into multi-q (typically triple-q) states that serve as a precursor to hexagonal skyrmion lattices.

Surprisingly, a small family of rare earth-based square-net compounds has recently been shown to host quadratic skyrmion lattices formed from double-q states \cite{moya2022incommensurate,moya2023real,neubauer2025correlation,yasui2020imaging,khanh2020nanometric,takagi2022square}. Square lattices are uniquely suited for exploring emergent magnetism. Their geometric simplicity, well-defined symmetry, and strong tunability enable an exceptional degree of microscopic control \cite{klemenz2019topological}. However, despite this versatility, all topological textures in these systems remain inherently 2D. Confining this hierarchy of textures to one-dimensional (1D) topological solitons, such as domain wall solitons \cite{kosevich1990magnetic,yoshimura2016soliton,nitta2022relations}, is paramount to understanding the properties of solitons across all dimensionalities. 

Unlike skyrmions, which typically have micromagnetic size and low-power mobility \cite{fert2013skyrmions, fert2017magnetic, marrows2021perspective} in 2D systems, domain walls act as coherent 1D soliton trains whose positions can be controlled deterministically by modest fields or currents. This deterministic motion forms the basis of racetrack memory concepts \cite{parkin2008magnetic}, where information is encoded in the sequence and spacing of domain walls. Shrinking these walls to the atomic scale maximizes storage density and minimizes the energy cost of manipulation. Atomically-sharp magnetic domain walls, only one or a few lattice spacings wide, are expected when Heisenberg-type exchange competes directly with magnetic anisotropy. Although they have been observed in some systems, including at structural phase boundary defects in antiferromagnets~\cite{krizek2022atomically}, monolayer-thick ferromagnetic nanostructures \cite{Pratzer2001DW_FeW110}, or in multiferroics at abrupt ferroelectric domain walls \cite{catalan2012domain}, in general the exchange interaction is an order if not orders of magnitude larger than the magnetic anisotropy, and atomically-sharp domain walls or field-tunable domain wall solitons remained elusive. 

In this work, we show that the square-net rare earth magnet EuRhAl$_{4}$Si$_{2}$ realizes the 1D topological limit. The system resides in a regime of highly frustrated magnetism, where competing ferro- and antiferro-magnetic RKKY exchange favors a Yoshimori spin spiral with three-site periodicity that competes directly with uniaxial magnetic anisotropy. The defining signature of this competition is a 1D collinear ferrimagnetic three-atom periodic up-up-down ($\uparrow \uparrow \downarrow$) state, evidenced by a robust $1/3$ magnetization plateau, as well as neutron diffraction. This state remains stable over a wide field range, within which atomically-sharp domain wall excitations emerge as single spin reversals along the chain, manifested as fine steps in magnetization and a stripe-like texture visualized by magnetic force microscopy (MFM). These excitations constitute 1D magnetic solitons. Their origin is captured by an effective $J_{1}$--$J_{2}$--$K$ model derived from density functional theory (DFT) and Monte Carlo simulations, which identify the soliton defect as the fundamental field-driven excitation in EuRhAl$_{4}$Si$_{2}$. These solitons are mobile, field-tunable, and embedded in a clean metallic environment, establishing this system as a unique platform for soliton-based racetrack elements operating at the ultimate spatial limit.

\section*{Signatures of magnetic soliton}\label{sec2}

\begin{figure}[h]
\centering
\includegraphics[width=0.9\textwidth]{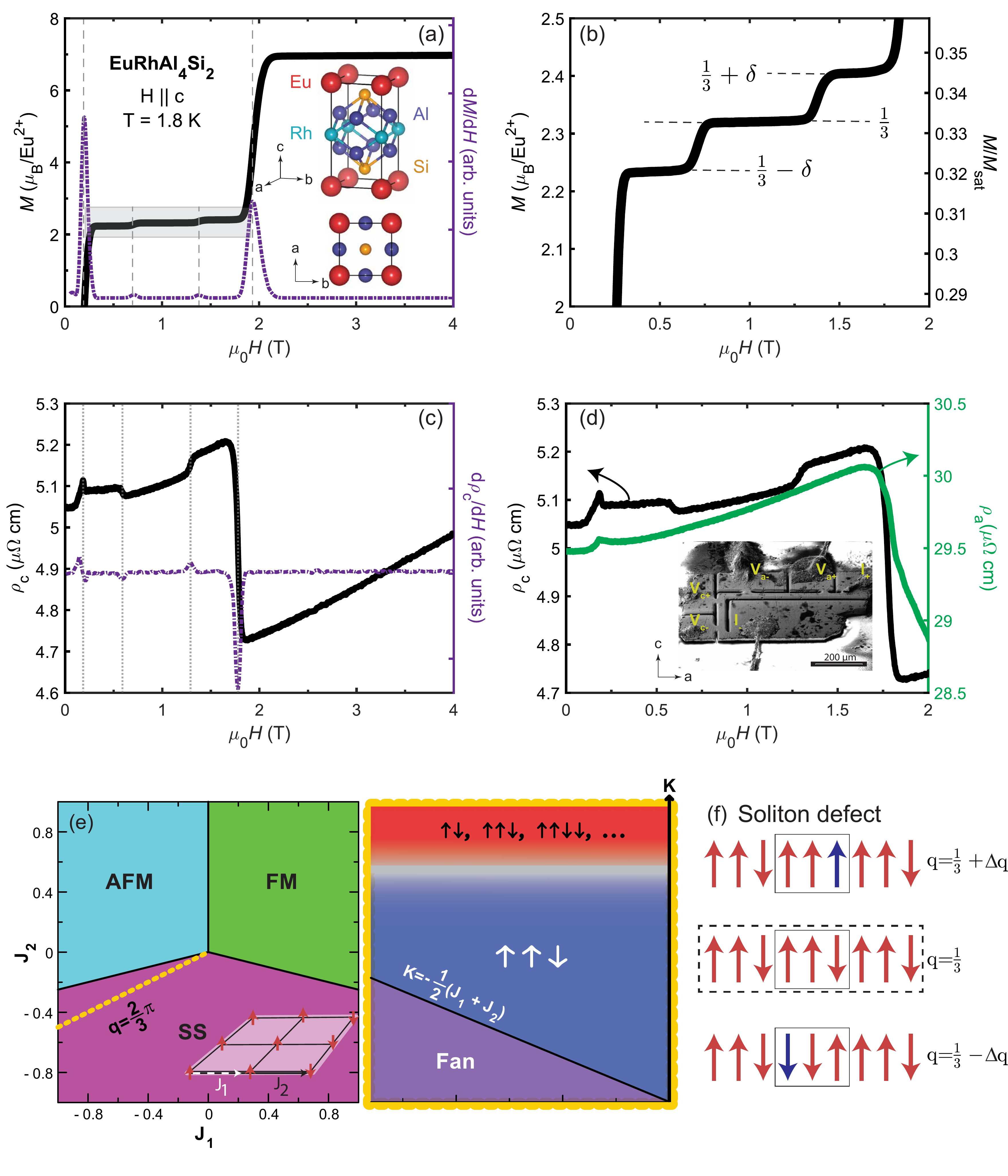}
\caption{\textbf{\mbox{Magnetization and soliton defects in EuRhAl$_4$Si$_{2}$.}} (a) Magnetization isotherm for EuRhAl$_4$Si$_{2}$ taken at T = $1.8$~K and $H \parallel c$  and the corresponding differential magnetic susceptibility $\text{d}M/\text{d}H$ (right axis, dashed line). The gray box indicates the region of interest. Inset shows the side and top view of the EuRhAl$_4$Si$_{2}$ crystal structure. (b) Enlarged view of the $M = M_\text{sat}/3 \cdot (1 \pm 1/100)$ plateaus and the corresponding spin arrangement of the Eu moments (inset). (c) Resistivity $\rho_c$ measured for the FIB-fabricated microstructure measured at T = $1.8$~K (left solid black line) and the corresponding derivative d\textit{$\rho_c$}/d\textit{H} (right axis, dashed line). (d) Resistivity for $\rho_c$ (black line) and $\rho_a$ (blue line). The inset shows the FIB-fabricated microstructured device on EuRhAl$_4$Si$_{2}$ single crystal. (e) Phase-diagram of the 1D frustrated $J_{1}$-$J_{2}$-$K$ Heisenberg model (inset effective lattice model). At the isotropic case (magnetic anisotropy $K=0$, left panel), there are three phases: ferromagnetic (FM), antiferromagnetic (AFM) and the non-collinear spin-spiral (SS) phase. The orange dotted line corresponds to the period-3 SS phase. Shown in the orange box (right panel) is the phase change along the dotted line as function of $K$, with the Fan state, up-up-down state up to the strong anisotropic limit. (f) Schematic of the soliton defect, the structure is not to scale. Bottom step: Time-reversal domains with domain wall soliton consisting of down-down-up spins (below). Middle step: 1D representation of the up-up-down order with q = $1/3$, Top step: additional formation of soliton defects (above).}\label{fig1}
\end{figure}

 EuRhAl$_4$Si$_{2}$ is a square-net compound that crystallizes in the BaMg$_{4}$Si$_{3}$-type structure (space group $P4/mmm$ \#123) \cite{maurya2014synthesis, maurya2015magnetic} (inset, Fig. \ref{fig1}a). Its magnetism originates in the localized spin-only Eu$^{2+}$ ions ($L = 0$), yet despite the imposed tetragonal symmetry and the lack of single-ion anisotropy, magnetization \textit{M} measurements as a function of  external magnetic field $H$ applied along the $c$ axis (Fig. \ref{fig1}a) reveal a $1/3$ magnetization plateau. The step, which occurs at $1/3$ \textit{M}$_\text{sat}$, persists up to $\sim 2\,$T before transitioning to the spin polarized (SP) state at saturation (\textit{M}$_\text{sat} =7\, \mu_{B}/$ Eu$^{2+}$). Two additional steps are observed around the $1/3$ plateau (gray box, Fig. \ref{fig1}a), seen more clearly in the derivative d\textit{M}/d\textit{H}. The additional steps correspond to values of $M = M_\text{sat}/[3(1 \pm \delta)]$, with $\delta \sim 1/100 \ll 1$ (Fig.~\ref{fig1}b). The two small steps indicate that the field evolution cannot be fully understood from frustrated exchange and magnetic anisotropy alone. Instead, they point to discrete spin-flip excitations that separate commensurate domains. In this picture, we demonstrate below that each additional step corresponds to the entry or rearrangement of a single soliton wall within an otherwise ordered $\uparrow \uparrow \downarrow$ background. 
 
In electrical transport measurements performed on a FIB-fabricated microstructure (Fig.~\ref{fig1}c-d) we observe the magnetic soliton response. For H $\parallel c$, $\rho_{c}$ tracks the magnetization field dependence. The central $1/3$ plateau exhibits the lowest resistivity, consistent with a commensurate state. The adjacent small plateaus show enhanced resistivity associated with the additional scattering from soliton walls. This directly links the transport anomaly to an electrical readout of the soliton lattice. In contrast, the in-plane resistivity $\rho_{a}$ displays the major transitions (low-field reorientation and spin flip) but lacks the fine steps, indicating that the conduction electrons are far less sensitive to soliton modulation in the basal plane. This anisotropy arises naturally if the domain walls are aligned along the in-plane direction while the spin modulation is along the $c$ axis propagation direction. 

DFT calculations indicate a spin-spiral ground state near $q=1/3$, as discussed in detail below, suggesting that frustrated exchange interactions naturally favor incommensurate modulations. Thermodynamic measurements (Supplementary Note 4), which confirm the ordering temperature $T_{N} \sim 11.6\,$K, in addition to the strong field dependence of the magnetic structure, imply that the effective Heisenberg exchange $J$ is similar in magnitude to the magnetic anisotropy. In this regime, the external magnetic field can not only reorient spins but fundamentally reshape the spin texture. As the anisotropy crosses a critical threshold, the system transitions from an incommensurate spin-spiral to a commensurate collinear out of plane stripe-like $\uparrow \uparrow \downarrow$ phase, as depicted by the single domain in Fig. \ref{fig1}e. This is described by the 1D Heisenberg model with competing nearest and next-nearest neighbor interactions $J_1$ and $J_2$, and competing magnetic anisotropy $K$ (Supplementary Note 2). Simultaneously, the $\uparrow \uparrow \downarrow$ state, being ferrimagnetic and non-compensated, produces a net moment and generates dipolar interactions that stabilize the magnetic domains. Together, these effects justify reducing the system to an effective 1D Heisenberg model, where domain wall solitons emerge as the fundamental excitations, tunable by both field and temperature. Fig. \ref{fig1}f illustrates the domain wall magnetic solitons corresponding to each of the steps. The bottom step ($1/[3 - \delta$]) comes as a result of the time-reversal domains with a domain wall soliton (boxed spins). The middle step is the commensurate $q = 1/3$ state where all spins have taken the $\uparrow \uparrow \downarrow$ configuration. The top step ($1/[3 + \delta$]) is reached with an additional soliton defect that flips spins into a new lowest energy state. 

\begin{figure}[h]
\centering
\includegraphics[width=1\textwidth]{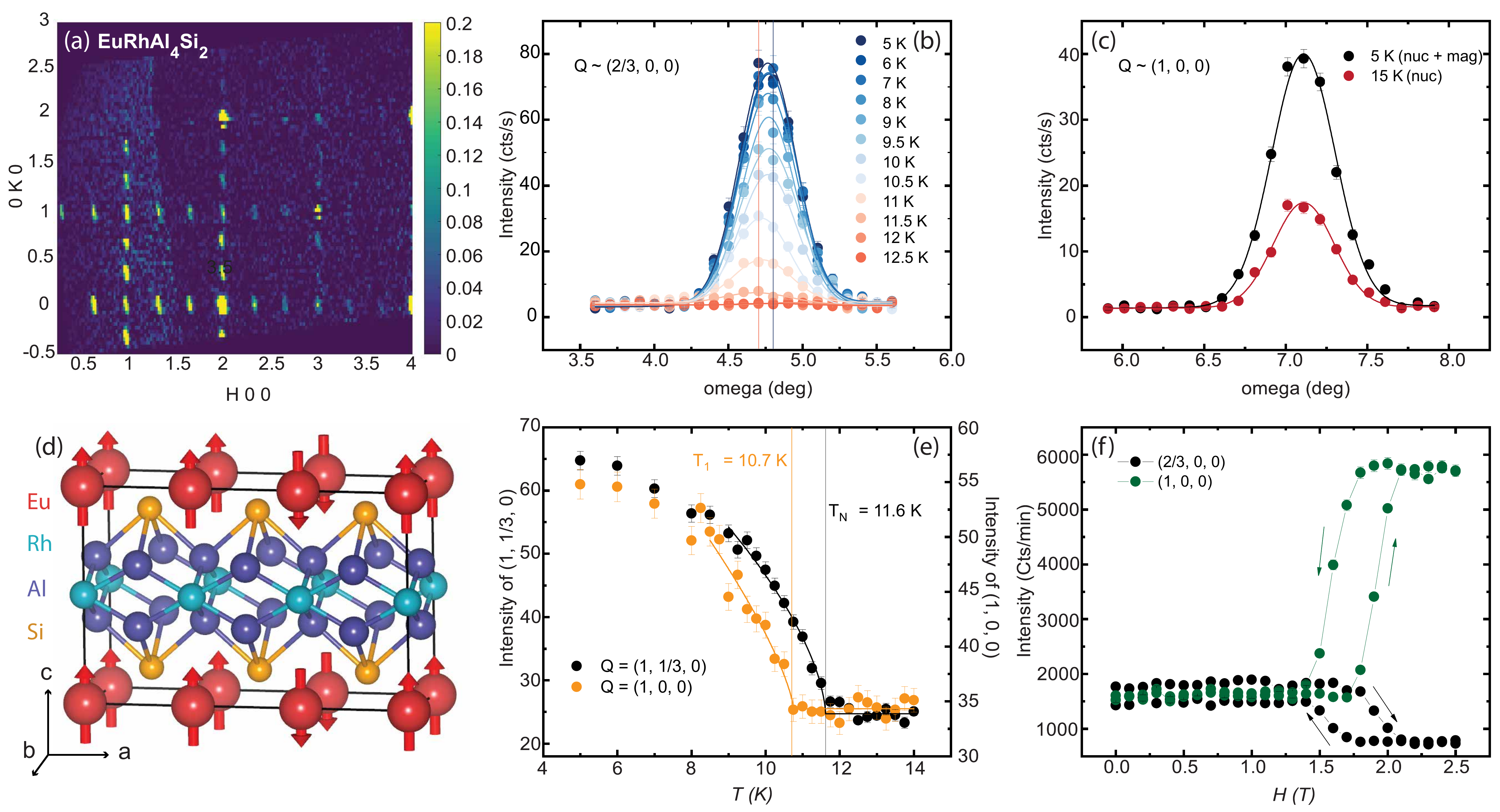}
\caption{\textbf{Up-up-down magnetic structure of \mbox{EuRhAl$_{4}$Si$_{2}$}}(a) Neutron diffraction patterns in the (HK0) scattering plane measured at $5$~K and $0$~T. Magnetic peaks are found at $\mathbf{k} = 0$, $\mathbf{k} = (1/3, 0 ,0)$ and $\mathbf{k} = (0, 1/3, 0)$. The latter two wave vectors are results of a- and b- direction magnetic twin domains. (b) Rocking curves of $(2/3, 0, 0)$ magnetic peak at zero field. Colors (blue to red) indicate temperatures from $5\,K$ to $12.5$~K. (c) Rocking curves of $(1, 0 ,0)$ Bragg peak measured at zero field. The intensity at $15$~K is from nuclear scattering, and the intensity at $5\,K$ is from both nuclear and magnetic scattering. (d) Up-up-down magnetic structure of EuRhAl$_{4}$Si$_{2}$ in the $3\!\times\! 1\times\! 1$ magnetic unit cell. (e) The temperature dependence of $(1, 1/3, 0)$ and $(1, 0 ,0)$ magnetic peaks at zero field. The solid lines represent critical exponent fitting. (f) The field hysteresis of $(2/3, 0 ,0)$ and $(1, 0 ,0)$ magnetic peaks measured at $1.5$~K. The intensity difference of $(2/3, 0 ,0)$ before and after magnetic field history shows magnetic twin-domain redistribution.}\label{fig2}
\end{figure}

\section*{Magnetic structure}\label{sec3}
To establish the microscopic origin of the soliton defects inferred from magnetization and transport and verify the effective 1D model, we determined the zero-field and in-field magnetic structure of EuRhAl$_4$Si$_{2}$ single crystals using neutron diffraction at $5\,$K. The refined lattice parameters at $5\,$K are $a = 4.202(4)\,$\AA\, and $c = 8.344(4)\,$\AA. Three different magnetic propagation vectors identify the magnetic peaks: $\mathbf{k}_1 = 0$, $\mathbf{k}_2 = (1/3, 0, 0)$, and $\mathbf{k}_3 = (0, 1/3, 0)$ (Fig. \ref{fig2}a-c). $\mathbf{k}_2$ and $\mathbf{k}_3$ indicate the AFM component with twinned magnetic domains as two-$q$ magnetic order was excluded due to unrealistic magnetic structure with a reduced magnetic moment. The coexisting $\mathbf{k}_1 = 0$ indicates that the magnetic structure is likely a square-wave rather than a sinusoidal wave with a single $\mathbf{k}_2$ or $\mathbf{k}_3$ in each magnetic domain. Furthermore, out-of-plane coupling is FM. The magnetic structure was solved by refining both nuclear and magnetic structures using over 70 magnetic and nuclear Bragg peaks. Symmetry analysis yields eight maximum magnetic space groups ( Supplementary Table 4, Note 5), of which Pm'm'm (\#47.252) best describes the measured magnetic peak intensities. The resultant magnetic structure is a twinned $\uparrow \uparrow \downarrow$ ferrimagnetic structure, propagating along the a- (or b-) axis in a $3\!\times\! 1\times\! 1$ (or $1\!\times\! 3\times\! 1$) magnetic supercell (Fig.~\ref{fig2}d).

The ordered magnetic moments, $m = 6.7 (1)\, \mu_\text{B}/$Eu, align along the c axis. This result is consistent with the measured moment from X-ray magnetic circular dichroism (XMCD) at the Eu $M_{4,5}$ edges (Supplementary Note 6) and the $1/3$ magnetization plateau with net magnetization $\sim 2.3\, \mu_\text{B}/$Eu. Magnetic domain fractions are refined to be $61.6 \%$ for $\mathbf{k}\,||\,a$ and $38.4 \%$ for $\mathbf{k}\,||\,b$. In a square lattice, both x- and y- directions are equal, which allows the domain wall to propagate along either crystallographic direction without geometric frustration. The order parameters were measured at zero field on the $(1, 0 ,0)$ and the $(1, 1/3, 0)$ magnetic peaks (Fig. \ref{fig2}e). We found two magnetic transition temperatures: $T_{1} = 10.7\,$K for the ferromagnetic component and $T_{N} = 11.6\,$K for the antiferromagnetic component in agreement with our thermodynamic and electrical transport properties (Supplementary Note 4,5). The magnetic field effect was measured using the same sample at $1.5\,$K and up to $2\,$T. Fig. \ref{fig2}f illustrates the magnetic field response of $(2/3, 0, 0)$ and $(1, 0, 0)$ magnetic peaks. By increasing the field to $2\,$T, the antiferromagnetic $(2/3, 0, 0)$ peak disappears, and the ferromagnetic $(1, 0, 0)$ peak becomes enhanced, which indicates a ferrimagnetic to ferromagnetic phase transition. This change marks the annihilation of the antiferromagnetic component between soliton walls as the lattice becomes increasingly ferromagnetic. As the magnetic field decreases, we observe hysteresis in both the $(2/3, 0, 0)$ and $(1, 0, 0)$ magnetic peaks, which aligns with the M-H curve. The hysteresis reflects the rearrangement of the twin domains as soliton walls change into the $\uparrow \uparrow \downarrow$ structure. It is noted that the intensity of $(2/3, 0, 0)$ becomes different after field history; this is understood by the redistribution of magnetic twin domains at the ferromagnetic-to-ferrimagnetic phase transition.  

\begin{figure}[h]
\centering
\includegraphics[width=0.9\textwidth]{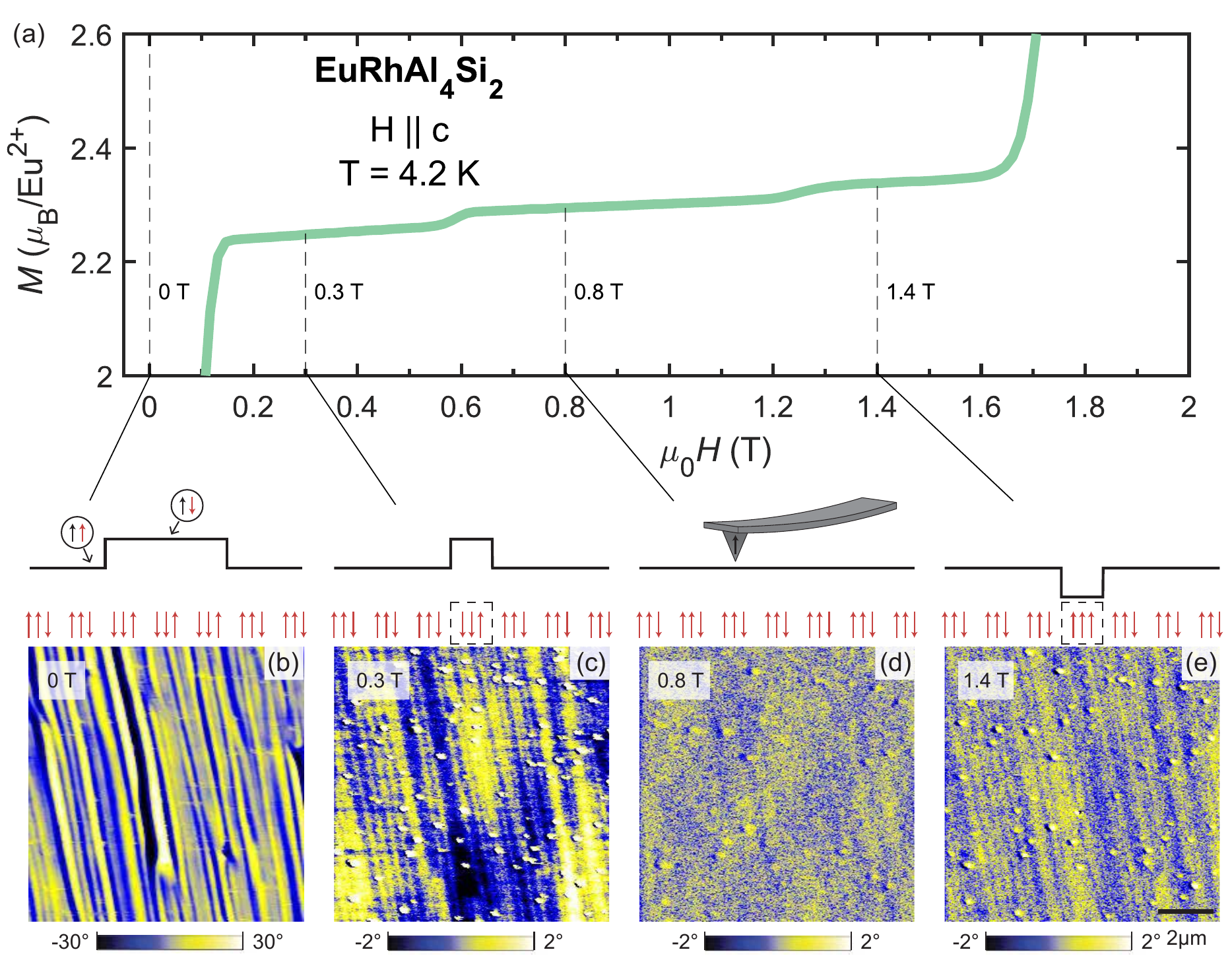}
\caption{\textbf{Magnetic field dependent MFM images of EuRhAl$_{4}$Si$_{2}$ at $4.2\,$K in the magnetically ordered state.} (a) Magnetization M as a function of magnetic field $\mu_{0}H$ (H $\parallel$ c. T = $4.2\,$K). (b) $0\,$T MFM image displaying the stripe-like $\uparrow \uparrow \downarrow$ and $\downarrow \downarrow \uparrow$ domains. The net magnetization and the  MFM phase contrast switch as it crosses from one domain (blue) to the next (yellow). (c) $0.3\,$T (bottom step, inset (c) MFM image, the majority of the spins are $\uparrow \uparrow \downarrow$ with a smaller phase contrast from the domain wall soliton $\downarrow \downarrow \uparrow$ as shown by the dashed boxed spins. (d) At the middle plateau (inset), $0.8\,$T, the MFM contrast disappears. This state corresponds to a uniform $\uparrow \uparrow \downarrow$ domain. (e) At $1.4\,$T, the top step, the periodic weak stripes reappear with opposite contrast from (c). Most spins are in the $\uparrow \uparrow \downarrow$-state with the domain wall soliton formed by $\uparrow \uparrow \uparrow$. Scanned areas are $10\mu$m$\times10\mu$m. Black lines illustrate the schematic MFM phase contrast and red arrows show their corresponding spin configurations.}\label{fig3}
\end{figure}

\section*{Visualization of domain walls}\label{sec4}
Next, we directly visualize the magnetic solitons inferred from magnetization, neutron diffraction and our 1D model using MFM. MFM measurements were performed on EuRhAl$_{4}$Si$_{2}$ under applied magnetic field at $4.2\,$K. The corresponding magnetization measurement is shown in Fig. \ref{fig3}a. MFM provides a real-space map of the out-of-plane stray field, allowing domain walls and soliton defects, \textit{i.e.} regions where the local magnetization reverses over a single lattice spacing, to be imaged through their characteristic phase contrast. The regions in which the magnetization of the sample is parallel (antiparallel) to the magnetized tip appear as negative (positive) phase shifts, allowing the distinction of $\uparrow \uparrow \downarrow$ and $\downarrow \downarrow \uparrow$ domains directly (Fig. \ref{fig3}b-e). At H = 0 (Fig. \ref{fig3}b), EuRhAl$_{4}$Si$_{2}$ exhibits stripe-like magnetic domains with widths of $\sim 200-400\,$nm and strong MFM contrast. The stripes run along either the $a$ or $b$ crystallographic axes, and their orientation rotates by $90^{\circ}$ in different regions of the same crystal (Supplementary Note 7). This is consistent with the crystallographic $C_{4}$ symmetry and the existence of the orientational magnetic twin domains. 

Upon increasing the field into the first incommensurate plateau at $0.3\,$T (Fig. \ref{fig3}c), the MFM image changes qualitatively. The wide zero-field domains collapse, and a new set of narrow stripes emerges, with much weaker phase contrast reflecting a decrease in the $\downarrow \downarrow \uparrow$ domains. These residual minority regions are confined to atomically-sharp boundaries, forming a 1D lattice of magnetic solitons separating extended $\uparrow \uparrow \downarrow$ segments. This configuration of spins explains the appearance of a magnetization plateau with values slightly smaller than $1/3$. At the center of the $1/3$ plateau (middle plateau Fig. \ref{fig3}a), the MFM contrast vanishes, consistent with a fully commensurate $\uparrow \uparrow \downarrow$ state in which no soliton walls remain. When the field is increased further into the upper incommensurate plateau, periodic stripes reappear but with opposite sign of the MFM contrast compared to the lower plateau (Fig. \ref{fig3}c). This reversal is associated with the soliton defect consisting of a single flipped spin within a three up spin background, yielding a magnetization value $M = M_{sat}/[3 (1 + 1/100)]$. The change in phase contrast suggests that the soliton lattice switches from a minority down to  minority up defects as the field increases through the $1/3$ plateau. Finally, once the system enters the spin polarized state (Supplementary Material), all stripes disappear, reflecting the absence of any soliton walls.
 
\begin{figure}[h]
\centering
\includegraphics[width=0.9\textwidth]{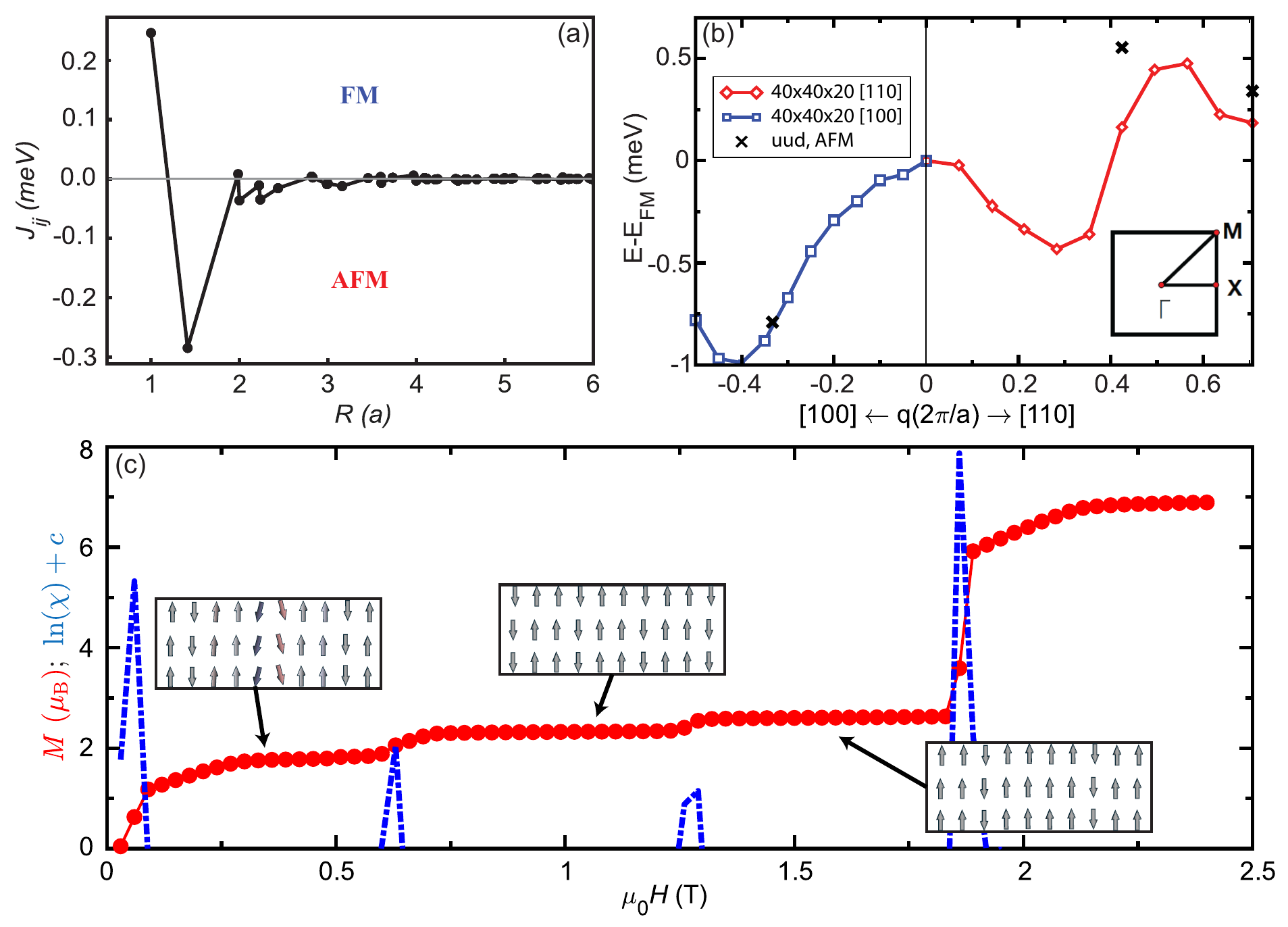}\caption{\textbf{Magnetic interactions and atomistic spin dynamics.}
(a) Real-space isotropic exchange interactions ($J_{ij}$) among the Eu magnetic atoms as function of the interatomic distance $R_{ij}$ given in units of the in-plane lattice constant $a$. (b) DFT spin-spiral calculations of EuRhAl$_{4}$Si$_{2}$ for $q$ along the X-$\Gamma$ (blue) and $\Gamma$-M (red) directions. Total energies for colinear (\ud, \uud) states along the two directions are indicated with crosses. (c) Magnetic moment per atom averaged over 3-atom unit cell, $M$ (red), and susceptibility $\chi$ (blue), obtained from a Monte Carlo simulation at $T = 0.1\,$K for a quasi-one-dimensional system with exchange parameters as described in the text. Shown is the transition from a spin-spiral ($H=0$, $M =0$) via a \uud~state ($M=7/3$) to a FM state ($M=7$) as function of magnetic field. The small steps indicate the formation of magnetic defects as indicated by the spin configurations.}\label{fig4}
\end{figure}

\section*{Magnetic interactions and simulations}\label{sec4}
We elucidate the magnetic interactions that stabilize the ground state by combining first-principles calculations with classical spin modeling, supported by photoemission experiments. DFT confirms a large Eu $4f$ moment close to $7\,\mu_B$, in agreement with the experiment, while induced moments on Rh, Al and Si remain below 0.5\% of the Eu moment with weak orbital overlap (Supplementary Table~1, Supplementary Note 1, Supplementary Note 8). Electronic correlations were improved using a Hubbard $U$ chosen to reproduce the Eu $4f$ peak observed by angle-resolved photoemission spectroscopy (ARPES; Supplementary Note~1).  Consistent with ARPES, the calculated electronic structure reveals only a few bands crossing the Fermi level, resulting in an almost semimetallic electronic structure.

The isotropic atomistic Heisenberg exchange parameters $J_{ij}$, calculated as a function of interatomic distance $R_{ij}$ between atoms at sites $i$ and $j$ (Fig.~\ref{fig4}a), display oscillatory ferro- and antiferro-magnetic pair interactions with a decaying amplitude, characteristic of RKKY exchange, as found in other square-net Eu and Gd compounds~\cite{Bouaziz2022,Miao2024}. The exchange interactions are weak, accounting for the low ordering temperature, and short ranged, extending up to three lattice constants, reflecting the semimetallic electronic structure~\cite{Bouaziz2024}. A detailed analysis (Supplementary Table~2) reveals weak ferromagnetic interlayer coupling along the $c$ axis and competing nearest- (nn) and next-nearest-neighbor (nnn) interactions $J_{ij}$ within the Eu planes, consistent with the nature of the compound comprised of Eu layers. 

Starting from a 2D Heisenberg model, we derive the $T=0$ phase diagram and show that the RKKY interaction network can be reduced to an effective 1D $J_1$--$J_2$ model of nn and nnn interactions along the square-net edges (inset of Fig.1e; Supplementary Note~2). The calculated exchange parameters, $J_1=-0.5$, $J_2=-0.14$~meV, place EuRhAl$_4$Si$_2$ at the phase boundary between the $p(2\times1)$ antiferromagnetic state (crystallographically equivalent to $p(1\times2)$) and an incommensurate spin-spiral phase (Fig.~\ref{fig1}e). Refinement using spin-spiral DFT calculations of the total energy $E(\mathbf q)$ for wave vectors along the in-plane high-symmetry directions $[100]$ and $[110]$ (Fig.~\ref{fig4}b) reveal a global minimum at $\mathbf q=(\pm 0.4,0,0)\,2\pi/a$, corresponding to a period of 2.5 lattice constants. The energy variations are between $-1$ and $0.5$~meV around the ferromagnetic state, and thus very small, indicating strong magnetic frustration and as typical for frustrated systems several nearly degenerate states, including the $p(2\times1)$ antiferromagnet and a period-3 spin spiral at $J_2/J_1=1/2$ (crosses in Fig.~\ref{fig4}b; Supplementary Table~3).

Although square-net rare earth compounds near this regime often stabilize skyrmion lattices under applied magnetic fields, surprisingly, the formation of a double-$q$ state is suppressed here by a large uniaxial anisotropy, calculated as $K=0.45$~meV for the $\uparrow\uparrow\downarrow$ configuration. This anisotropy competes directly with the effective exchange interactions and stabilizes the experimentally observed $\uparrow\uparrow\downarrow$ out of plane state as the lowest-energy configuration. The resulting interplay of competing exchange interactions at $J_2/J_1=1/2$ and sizable magnetic anisotropy places EuRhAl$_4$Si$_2$ in an unconventional regime of the extended $J_1$--$J_2$--$K$ Heisenberg model. In this framework, magnetic anisotropy interpolates between the isotropic Heisenberg limit ($K=0$) and the  axial next-nearest-neighboring Ising (ANNNI) model~\cite{Bak:1982_Rev}, which at $J_2/J_1=1/2$ exhibits a multiphase boundary between period-2 and period-4 antiferromagnetic order. Our analysis of the period-3 spin-spiral phase as a function of increasing anisotropy (Fig.~\ref{fig1}e; Supplementary Note~2) shows a continuous evolution from a spiral to a fan (Y-type) state, with the intersite angle progressively deviating from $120^\circ$, before transitioning to the collinear $\uparrow\uparrow\downarrow$ phase for $K > -(J_1+J_2)/2$ with $J_1,J_2<0$.

To understand the formation of solitonic defects as a function of an applied magnetic field in a frustrated magnet with many nearly degenerate magnetic phases, we performed Monte Carlo calculations at temperatures between 0.1 and 1.5 K that are consistent with our set of parameters $J$ and $K$. At zero magnetic field, the energy landscape exhibits many local minima due to  degenerate and nearly-degenerate states ($\uparrow\uparrow\downarrow$, $\downarrow\downarrow\uparrow$,  $\downarrow\uparrow$). In Fig.~\ref{fig4}c, which qualitatively reproduces the experimental magnetic field dependence $M(H)$, we see that as the magnetic field increases, the population of the $\uparrow\uparrow\downarrow$ state increases, including the remaining defects. As the magnetic field increases, the remaining $\downarrow\uparrow\downarrow$ defects initially  disappear and new ones ($\uparrow\uparrow\uparrow$) are introduced  in the background of $(\uparrow\uparrow\downarrow)$ until the transition to the field-saturated ferromagnetic state. The agreement between experiment and simulation confirms that the magnetization plateaus directly signal a field-tuned domain wall soliton lattice. The small steps observed experimentally at incommensurate q values reflect soliton density variations as the field tunes the systems between commensurate and incommensurate phases.

\section*{Discussion}\label{sec12}
The present work demonstrates that EuRhAl$_{4}$Si$_{2}$ realizes an atomically sharp magnetic domain-wall soliton stabilized by the finely-tuned competition between frustrated exchange and magnetic anisotropy in a square-net lattice. This combination is unusual as in most centrosymmetric rare earth systems, competing RKKY exchange selects either a commensurate collinear structure or an incommensurate spiral, but rarely a regime where both states are nearly degenerate. In EuRhAl$_{4}$Si$_{2}$, this energetic proximity produces a commensurate-incommensurate boundary where the lowest energy excitations are single spin flip soliton defects on a single Eu site. The square-net geometry, with equivalent $a$ and $b$ directions and naturally twinned magnetic domains, provides an ideal platform for solitons to propagate and reorient without crystallographic frustration. This enables the realization of a mobile soliton that is directly accessible both to real-space imaging and to bulk transport and magnetization measurements.

This mechanism differs fundamentally from the origin of the devil's staircase \cite{bak1986devil} in other magnetic materials. In quantum spin systems, the steps originate from energy gaps in the many-body excitation spectrum under the commensurability condition n(S-m) = integer \cite{oshikawa1997magnetization}. This behavior is characteristic of low-spin quantum systems, where fluctuations stabilize gapped ground states \cite{kageyama1999exact,shiramura1998magnetization}. In CeSb \cite{rossat1977phase,von1979devil} and related rare earth monopnictides \cite{singha2021evolving, chen2017possible}, the staircase reflects a continuous spin-density modulation whose ordering wavevector locks into rational fractions under field and temperature tuning. In contrast, EuRhAl$_{4}$Si$_{2}$ host classical localized Eu$^{2+}$ moments in which the fine structure around the $1/3$ plateau originates from the insertion of real-space solitons at the boundary between the commensurate $\uparrow \uparrow \downarrow$ order and the nearby incommensurate spiral. The absence of significant induced moments on the nearby site (Si/Rh), the XMCD sum rules, and DFT calculations show that the magnetism is dominated by Eu$^{2+}$, ruling out competing itinerant mechanisms. Likewise, the small step height and its field evolution are incompatible with multi-$q$ spin-density waves or higher-harmonic lock-in transitions. Instead, the soliton density inferred from magnetization closely matches the soliton spacing directly observed by MFM. 

The soliton structure realized here also invites a conceptual analogy to the Su-Schrieffer-Heeger (SSH) model \cite{su1979solitons}. In the effective 1D chain defined by the $\uparrow \uparrow \downarrow$ pattern, the two consecutive up-spin bonds ($\uparrow \uparrow$) and the isolated down-spin bond ($\downarrow$) play the role of the alternating strong and weak links. A domain-wall soliton, created by shifting the phase of the $\uparrow \uparrow \downarrow$ pattern by one site, maps onto the topological midgap defect of an SSH chain. Within this picture, the field-induced soliton steps in the magnetization correspond to tuning the occupancy of these topological defects. Although the analogy is not fermionic, it highlights that the soliton in EuRhAl$_{4}$Si$_{2}$ results from topologically protected domain walls stabilized by a dimerized exchange-anisotropy landscape, and their motion corresponds to sliding a robust topological defect through a periodic background. The discovery of such atomically-sharp, field-tunable solitons in a rare earth square-net magnet opens an avenue towards cryogenic spintronic devices based on controllable magnetic domain walls. The soliton walls in EuRhAl$_{4}$Si$_{2}$ are narrow, mobile, and sensitive to modest magnetic fields, suggesting that current-driven motion, analogous to racetrack memory domain wall dynamics, may be achievable at low temperatures. Their topological robustness and the crystalline symmetry, which allows the soliton to propagate along either in-plane axis, make EuRhAl$_{4}$Si$_{2}$ an attractive platform for directional control or junction based soliton routing. Time-resolved imaging and transport experiments capable of tracking soliton drift and pinning collective dynamics are underway. 

In summary, by integrating magnetization, anisotropic magnetotransport, neutron diffraction, XMCD, MFM imaging, ARPES, DFT and classical spin modeling, we identify EuRhAl$_{4}$Si$_{2}$ as a model system in which frustrated exchange and magnetic anisotropy cooperate to stabilize a collinear $\uparrow \uparrow \downarrow$ ground state and an associated lattice of atomically-sharp magnetic solitons. The agreement between real-space imaging, magnetization, and theoretical modeling establishes a unified picture of the field-driven soliton physics. More broadly, our results demonstrate that, even in centrosymmetric magnets with purely isotropic exchange, sharp topological domain walls can emerge, opening a new route to tunable soliton-based functionalities in square-net materials.

\section*{Methods}\label{sec11}
\noindent\textbf{Sample growth and characterization.} Single crystals of EuRhAl$_{4}$Si$_{2}$ were grown in an excess of aluminum (Al) and silicon (Si) via a self-flux technique. Europium (Eu), rhodium (Rh), Al and Si were mixed in a ratio of 1:1:80:13.2 then placed in an alumna crucible and evacuated in a quartz ampule. The mix was heated to 1100$^{\circ}$C over 10 hours and subsequently slowly cooled over a period of 185 hours down to 750$^{\circ}$C, after which the crystals were separated from the excess liquid flux using a centrifuge. EuRhAl$_{4}$Si$_{2}$ forms as plate-like crystals with the biggest surface area corresponding to the crystallographic $ab$ plane. The largest crystals have lateral sizes up to 3 mm. The single crystals were confirmed to belong to the P4/mmm (space group \#123) tetragonal structure type  with powder X-ray diffraction. A Rietveld structural refinement was achieved and fit to the measured intensities. 

\noindent\textbf{Electrical transport and magnetization measurements.} 
The electrical transport experiments were carried out in a standard four-probe geometry in a Quantum Design (QD) Dynacool system, with field up to $14\,$T, applied current j = $150 \mu$A and frequency f = 12.2 Hz. Temperature- and magnetic-field dependent magnetization measurements were performed in a QD, Magnetic Property Measurement Systems (MPMS3) with vibrating sample magnetometer option (VSM). A single crystal of EuRhAl$_{4}$Si$_{2}$ was mechanically thinned to $\sim 20\,$ $\mu$m along the ac plane. The thinned sample was then transferred to a sapphire subtrate and electrical contacts were made with a two-component epoxy using $12.5\,$ $\mu$m wires. A microstructure in the form of a L-shaped device was patterned using a Ga-focused ion beam at $30\,$kV and $65\,$nA.

\noindent\textbf{Neutron diffraction experiments.} 
Neutron diffraction measurements were conducted at HB-3A DEMAND \cite{cao2018demand} at the High Flux Isotope Reactor at Oak Ridge National Laboratory. The experiment used a Si (331) monochromator with a wavelength of 1.003 \AA. In the zero-field experiment, the sample was mounted on the four-circle goniometer and cooled down to 5K using a helium closed-cycle refrigerator. In the field experiment, the sample was mounted on an aluminum sample holder and cooled down to 1.5K using a cryomagnet. The data reduction used a ReTIA package \cite{hao2023machine}. Adsorption correction used WinGX \cite{farrugia2012wingx}. The symmetry analysis used Bilbao Crystallography Server \cite{aroyo2006bilbao}. The structure refinement used Fullprof \cite{rodriguez1993recent}.

\noindent\textbf{Magnetic Force Microscopy.} 
As-grown surfaces of EuRhAl$_4$Si$2$ crystals were scanned using a temperature-variable AFM system (Attocube) in a dual pass mode (lift height $\sim 35\,$nm) with commercial Co/Cr-coated tips for MFM imaging.

\noindent\textbf{X-ray magnetic circular dichroism.} 
XMCD measurements were conducted using the VEKMAG instrument at BESSY-II in Berlin, Germany \cite{noll2016mechanics}. A single crystal of EuRhAl$_4$Si$2$ was cleaved in situ to prevent surface oxidation. The XMCD measurements were performed in total electron yield (TEY) detection mode, utilising 77\% circularly polarized soft X-rays at the Eu $M_{4,5}$ edges. Magnetic fields ranging from -3 T to +3 T were applied parallel to both the X-ray beam and the c axis of the sample.

\noindent\textbf{DFT--Density-functional calculations.} 
The simulations were performed using the all-electron full-potential Korringa–Kohn–Rostoker (KKR) Green function method implemented in the {\sc juKKR} code~\cite{russmann6judft}. The scalar-relativistic approximation was employed extended by including spin–orbit coupling, which was treated self-consistently.  The generalized gradient approximation (GGA) was used for the exchange–correlation potential~\cite{Perdew1996}. The Eu $4f$-orbitals were treated within Hubbard $U$-corrected density functional theory (DFT+$U$)~\cite{Dudarev2019}, with a Hubbard $U$ of $8.5$~eV to reproduce the position of the Eu $4f$-peak observed in photoemission experiments (Supplementary Note 1). The spherical harmonic expansion of the Green function was truncated at an angular momentum cutoff of $\ell_{\text{max}} = 4$. The integration of the Green function for achieving self-consistency included an energy contour of $51$ points, a dense $k$-mesh in the Brillouin zone of $30 \times 30 \times 30$ and a smearing temperature of $500$~K. Magnetic interactions were extracted using the infinitesimal rotation method~\cite{Szilva2023}, starting from a collinear ferromagnetic state using a lower smearing temperature of $400$~K and a denser $k$-mesh of $60 \times 60 \times 60$ points.
The {\sc Fleur} code~\cite{fleurCode}, which employs the full-potential linearized augmented plane-wave (FLAPW) method~\cite{Kurz:04}, was used to treat spin-spiral states in the chemical unit cell via the generalized Bloch theorem. Vector–spin density functional theory within the $U$-corrected local density approximation (LDA) ($U=6.7$~eV, $J=0.7$~eV)~\cite{VWN} was then applied to compute spin-spiral energies and magnetic ani\-sotropies.
Reciprocal space was sampled with $40 \times 40 \times 20$ $k$-points for the spin-spiral calculations and $12 \times 32 \times 16$  $k$-points for the magnetic anisotropy of the \uud~structure. All calculations are based on experimental structural data in agreement with ~\cite{maurya2014synthesis}.

\noindent\textbf{Atomistic spin dynamics}
These simulations were performed using the {\sc Spirit}-code~\cite{spiritCode,Mueller2019}, a framework for spin dynamics simulations. A quasi-1D geometry with $300 \times 8 \times 3$ sites in a tetragonal lattice was used with nearest-neighbor Heisenberg couplings $J_{1x}=-0.25$, $J_{1y}=0.25$, and $J_{1z}=0.008$, as well as a next-nearest neighbor coupling $J_{2x}=0.125$ and a uniaxial anisotropy $K_z = 0.125$ in relative units. The magnetic field, $B_z$, was ramped up continuously from zero to saturation of the magnetization. The Monte Carlo simulations were performed at $0.1 - 1.5$~K.

\noindent\textbf{Photoemission spectroscopy}
The photoemission spectroscopy measurements (see Supplementary Fig. 5) were carried out at the QMSC beamline of the Canadian Light Source using a R4000 electron analyzer with a horizontal slit. In the presented data, the $p$ polarized photons were aligned 45$^\circ$ relative to the $a$ axis of the crystal. The sample was cleaved in-situ in an ultrahigh vacuum better than $6\times10^{-11}$ Torr and measured at a temperature below 17 K. The detailed temperature dependent angle-resolved photoemission data will be presented in an upcoming work.

\backmatter



\bmhead{Acknowledgements}

The work at Rice University was primarily supported by the Robert A. Welch Foundation Grant No. C-2114 and the Vannevar Bush Faculty Fellowship ONR-VB N00014-24-1-2048. The work at Rutgers University was supported by the DOE under Grant No. DOE: DE-FG02-07ER46382. Y.H. and H.C. acknowledge support by the U.S. DOE, Office of Science, Office of Basic Energy Sciences, Early Career Research Program Award KC0402020, under Contract DE-AC05-00OR22725. This research used resources at the High Flux Isotope Reactor, a DOE Office of Science User Facility operated by the Oak Ridge National Laboratory. The beam time was allocated to HB-3A DEMAND on proposal number IPTS-31843. V.U., C.L., and F.R. acknowledge financial support for the VEKMAG project and for the PM2-VEKMAG beamline by the German Federal Ministry for Education and Research (BMBF 05K2010, 05K2013, 05K2016, 05K2019, 05K2022) and by HZB. S.M., E.M., Y.G., Y.Z. and M.Y. acknowledge partial support by the Department of Defense Air Force Office of Scientific Research under Grant No. FA9550-21-1-0343. The work at Los Alamos National Laboratory was carried out under the auspices of the US Department of Energy (DOE) National Nuclear Security Administration under Contract No. 89233218CNA000001, and was supported by the Los Alamos National Laboratory (LANL) LDRD Program, and in part by the Center for Integrated Nanotechnologies, an Office of Science User Facility operated by the U.S. Department of Energy (DOE) Office of Science, in partnership with the LANL Institutional Computing Program for computational resources.
S.B.\ acknowledges funding by the European Research Council grant 856538 (project "3D MAGIC"), the Deutsche Forschungsgemeinschaft (DFG, German Research Foundation) through SFB-1238 (project C1), and by the European CoE MaX ``Materials design at the Exascale'' (Grant No.\ 824143) funded by the EU. Part of the photoemission experiments were performed at the Canadian Light Source, a national research facility of the University of Saskatchewan, which is supported by Canada Foundation for Innovation (CFI), the Natural Sciences and Engineering Research Council of Canada (NSERC), the National Research Council (NRC), the Canadian Institutes of Health Research (CIHR), the Government of Saskatchewan, and the University of Saskatchewan.


\bmhead{Contributions}
We are thankful to Yoshinori Tokura and Steven Kivelson for fruitful discussions. E.M.\ conceived the project. K.A., Y.Z., S.M., S.B., and E.M.\ contributed to the analysis. K.A. synthesized the crystals and performed magnetotransport and magnetization measurements with help from S.M., Y.G.. S.M. fabricated the FIB device, J.B.\ and G.B.\ performed the DFT calculations. K.A., C.L., and J-X.Z.\ performed the orbital projection analysis, G.B.\ performed the atomistic spin dynamics. S.B.\ constructed the model. The ARPES work was done by Y.Z. and M.Y. The neutron experiments were done by Y.H., and H.C.. XMCD was measured by V.U., C.L., and F.R. MFM measurements were performed by K.D. and S-W.C. All authors discussed the results and contributed to the writing of the manuscript.
\section*{Declarations}

The authors declare no competing interests.

\bibliography{sn-bibliography}

\end{document}